\documentclass[fleqn,twoside]{article}
\usepackage{espcrc2}

\usepackage{graphicx}
\usepackage[figuresright]{rotating}

\newcommand{\AmS}{{\protect\the\textfont2
  A\kern-.1667em\lower.5ex\hbox{M}\kern-.125emS}}

\title{$|V_{cb}|$ and $|V_{ub}|$ Determinations from Inclusive Semileptonic $B$
Decays}

\author{Changhao Jin\address[MCSD]{School of Physics, 
        University of Melbourne, 
        Victoria 3010, Australia}}
        
\begin{document}

\begin{abstract}
The determinations of $|V_{cb}|$ and $|V_{ub}|$ from inclusive
semileptonic $B$ decays are reviewed. A comparison of the light-cone
approach with the heavy quark expansion approach is made. The perspective for
further
improvement of the $|V_{cb}|$ and $|V_{ub}|$ determinations is discussed.
\vspace{1pc}
\end{abstract}

\maketitle

\section{INTRODUCTION}

$|V_{cb}|$ and $|V_{ub}|$ are two of the fundamental constants of the
Standard Model. They need to be determined by experimental measurements.
These can be done with leptonic, semileptonic, or nonleptonic decays of
$B$ mesons.
In this talk I will discuss the determinations of  $|V_{cb}|$ and
$|V_{ub}|$ from inclusive semileptonic B decays. An observable in these
processes is schematically related to $|V_{cb}|$ or $|V_{ub}|$ in the
form
\[
{\rm observable}=|V_{c(u)b}|^2\cdot T,
\]
where $T$ is a quantity derived from theory.  Theory may also be involved
in the experimental analysis to obtain the observable. With the
theoretical relationship between the observable and $|V_{cb}|$ or
$|V_{ub}|$, the measured quantity can be converted into a value of
$|V_{cb}|$ or $|V_{ub}|$.  The main obstacle in the theoretical
description of the weak decay processes is long-distance strong
interaction effects. There are two classes of observables: (1) routine
observables, such as branching fractions and lifetimes,
and (2) theory-motivated observables, such as the differential decay
rate $d\Gamma/dw$ at zero recoil ($w=1$) in the exclusive semileptonic decay 
$B\to D^*l\nu$ for determining $|V_{cb}|$.
The latter is motivated from theory because its relationship with
$|V_{cb}|$ or $|V_{ub}|$ is theoretically clean and model-independent in
some sense.

\section{THEORY OF INCLUSIVE SEMILEPTONIC $B$ DECAYS}

Strong interaction effects on inclusive semileptonic $B$ decays are
contained in the hadronic tensor
\begin{eqnarray}
W_{\mu\nu}&=&-\frac{1}{2\pi}\int d^4y e^{iq\cdot y}
\langle B\left|[j_{\mu}(y),j^{\dagger}_{\nu}(0)]\right|B\rangle \nonumber\\
 &=&-g_{\mu\nu}W_1 + \frac{P_{\mu}P_{\nu}}{M_B^2} W_2 
-i\varepsilon_{\mu\nu\alpha\beta} \frac{P^{\alpha}q^{\beta}}{M_B^2}W_3
\nonumber\\
& &+ \frac{q_{\mu}q_{\nu}}{M_B^2} W_4 
 + \frac{P_{\mu}q_{\nu}+q_{\mu}P_{\nu}}{M_B^2} W_5.
\label{eq:comm2}
\end{eqnarray}
where the charged current 
$j_{\mu}(y) = \bar{q}(y)\gamma_{\mu}(1-\gamma_5)b(y)$ with $q=c, u$.
$P$ and $q$ denote the momenta of the $B$ meson and the virtual $W$ boson,
respectively.  
The central issue in the theory of inclusive semileptonic $B$
decays is to
compute the structure functions $W_{1-5}$, which are {\it a priori} 
independent. In the last several years, theoretical
tools have been developed to address the issue.

\subsection{Light-cone (LC) approach}

The starting point of the LC approach \cite{jp1,jin1,jp2,jin2,eur} to 
inclusive B decays is the
light-cone expansion. Since the $B$ meson is heavy, $M_B\gg\Lambda_{\rm
QCD}$, light-cone separations between the currents dominate the decay
dynamics. In analog to the analysis of deep inelastic scattering, the
light-cone expansion of the matrix element in the neighborhood of the
light-cone, $y^2=0$, provides a systematic way of calculating
nonperturbative QCD effects:
\begin{equation}
\langle B|[j^\mu(y), j^{\nu\dagger}(0)]|B\rangle= \sum_n 
C_n^{\mu\nu}(y, P)(y^2)^n.
\end{equation}

At leading twist, the structure functions are related to a single
universal distribution function:
\begin{eqnarray}
W_1 & = & 2[f(\xi_+) + f(\xi_-)]\, ,\label{eq:w1}\\
W_2 & = & \frac{8}{\xi_+ -\xi_-}[\xi_+f(\xi_+)-\xi_-f(\xi_-)]\,,\label{eq:w2}\\
W_3 & = & -\frac{4}{\xi_+-\xi_-} [f(\xi_+) -f(\xi_-)]\, ,\label{eq:w3}\\
W_4 & = & 0\, ,\label{eq:w4}\\
W_5 & = & W_3\, ,\label{eq:w5}
\end{eqnarray}
where $\xi_\pm = (q^0\pm\sqrt{|\vec{q}|^2+m_q^2})/M_B$. The distribution 
function is defined as
\begin{eqnarray}
f(\xi)&=&\frac{1}{4\pi}\int \frac{d(y\cdot P)}{y\cdot P}e^{i\xi y\cdot P}
\nonumber\\
& &\times\langle B|\bar{b}(0)y\cdot\gamma U(0,y)b(y)
|B\rangle |_{y^2=0}.
\label{eq:distr3}
\end{eqnarray}
It gives the probability of finding a $b$-quark with momentum $\xi P$ inside
the $B$ meson with momentum $P$ \cite{bareiss,jpp}.

The leading nonperturbative QCD effect is encoded in the distribution
function. What is
known about the distribution function?  First, the normalization of it is
exactly known to be $\int_0^1 d\xi f(\xi)=1$,
because of the $b$-quark number conservation in strong interactions. Second,
the gross shape of it is determined. In the free quark limit,
$f(\xi) = \delta(\xi-m_b/M_B)$. The mean value and variance of the
distribution function have been calculated using the heavy quark
effective theory. Consequently, the distribution function is known to be
sharply peaked around $m_b/M_B$ with a narrow width of order
$\Lambda_{\rm QCD}/M_B$.

\subsection{Comparison with the heavy quark expansion (HQE) approach}

Another approach to inclusive $B$ decays is the heavy quark expansion
approach \cite{chay,bigi1,bigi2,bigi3,manohar,blok,neubert,mannel}. 
The starting point of the HQE approach is the
local operator product expansion. It exploits the large b-quark mass to
perform a local operator product expansion ($y\to 0$) of the
time-ordered product of currents:
\begin{equation}
T[j^\mu(y),j^{\nu\dagger}(0)]=\sum_d C_d^{\mu\nu}(y) O_d(0).
\end{equation}

The starting point of the HQE approach is different from that of the LC
approach. The HQE approach is based on a short-distance ($y\to 0$) expansion
in local operators of increasing dimension, while the LC approach is
based on a non-local light-cone ($y^2\to 0$) expansion in matrix elements of 
increasing twist. The light-cone expansion includes not only local
contributions ($y=0$), but also non-local contributions
($y\neq 0$). Thus, the LC approach contains more dynamic effects.

In addition, the HQE approach assumes quark-hadron duality and has to use
the quark phase space, while the LC approach does not rely on
quark-hadron duality and uses the physical hadron phase space. The
singularities appear at the endpoints of the lepton energy spectra in
both $B\to X_u l\nu$ and $B\to X_c l\nu$ calculated in the HQE approach,
implying the breakdown of the operator product expansion near the
boundaries of phase space.  There are no endpoint singularities in the
lepton energy spectra calculated in the LC approach. 

The partial resummation of the heavy quark expansion introduces a different
distribution function (``shape function'') \cite{bigi3,neubert,mannel}.
It should be noted that the correction to the leading contribution given
in terms of the shape function in the HQE approach is of order
$\Lambda_{\rm QCD}/m_b$, while the correction to the leading contribution
given in terms of the distribution function in the LC approach is of
order $(\Lambda_{\rm QCD}/M_B)^2$.

\subsection{A manifestation of quark-hadron duality violation}

The physical phase space at the hadron level is larger than the phase
space at the quark level. Therefore, calculations assuming quark-hadron
duality cannot account for the rate due to the extension of phase space from 
the quark
level to the hadron level. This rate missing is a clear manifestation of
the violation of quark-hadron duality. This is the case for the HQE
approach.

Since the LC approach does not rely on quark-hadron duality and uses the
hadron phase space, a comparison of the decay rates calculated in the LC
and HQE approaches can quantitatively demonstrate how large duality is
violated. It has been found that the total rate for the inclusive
semileptonic decay $B\to X_c l\nu$ ($B\to X_u l\nu$)
calculated in the LC approach is about $14\%$ ($12\%$) larger than the
HQE approach \cite{jin1,pl}.
The nonperturbative QCD correction changes sign: It increases the total
rate in the LC approach, while it decreases the total rate in the HQE
approach. Therefore, there is significant duality violation in the HQE
approach. In particular, it cannot include the phase space effect.
From the measured inclusive semileptonic branching fractions and the lifetime
of the $B$ meson, the LC approach would give rise to smaller values of
$|V_{cb}|$ and $|V_{ub}|$ than the HQE approach, as I will discuss in more 
detail below.

\section{CONVENTIONAL METHODS}

Conventionally, the routine observables, the branching fractions and the
lifetime of the $B$ meson, have been used to determine $|V_{cb}|$ and
$|V_{ub}|$ from inclusive semileptonic $B$ decays. These observables are
related to $|V_{cb}|$ or $|V_{ub}|$ in the form
\begin{eqnarray}
{{\cal B}(B\to X_c l\nu)\over\tau_B} &=&|V_{cb}|^2\cdot\gamma_c , \\
{{\cal B}(B\to X_u l\nu)\over\tau_B} &=&|V_{ub}|^2\cdot\gamma_u.
\end{eqnarray}
Theory is needed to compute $\gamma_c$ and $\gamma_u$.
Besides parametric uncertainties, additional uncertainties in the
theoretical calculations of $\gamma_c$ and $\gamma_u$ stem from the
assumption of quark-hadron duality in the HQE approach and the 
detailed shape of the distribution function, which is unknown at present, 
in the LC approach.

The uncertainty due to the assumption of quark-hadron duality is
irreducible. The errors usually quoted in the determinations of
$|V_{cb}|$ and $|V_{ub}|$ from inclusive semileptonic $B$ decays using
the HQE calculations should be interpreted with caution. The significant
uncertainty from the assumption of quark-hadron duality has not been
included in the errors. In contrast, the uncertainty due to the shape of
the distribution function in the LC approach can be reduced. The
distribution function is universal: It incorporates bound state
effects in inclusive $B$ decays, 
including semileptonic \cite{jp1,jin1,jp2,jin2},
radiative \cite{eur}, and nonleptonic \cite{he,bsx} inclusive $B$ decays. 
Measurements of decay distributions and
moments in these processes will impose constraints on the distribution
function. Especially, the $\xi_u$ spectrum
[$\xi_u=(q^0+|\vec{q}|)/M_B$] in $B\to X_u l\nu$ and the photon energy
spectrum in $B\to X_s\gamma$ are directly proportional to the
distribution function. These spectra are most sensitive to the shape of
the distribution function. Measurements of these spectra may thus provide
the most stringent constraint on the distribution function. Moreover,
other nonperturbative QCD methods, such as lattice QCD, may help to
determine the shape of the distribution function, too.

Including the uncertainties from the input parameters and the shape of
the distribution function, the calculations in the LC approach 
yield \cite{jin1,pl}
\[
\gamma_c=49\pm 9 \ {\rm ps}^{-1}, \qquad 
\gamma_u=76\pm 16 \ {\rm ps}^{-1}.
\]
These results allow reliable determinations of $|V_{cb}|$ and $|V_{ub}|$
without the assumption of quark-hadron duality. The accuracies could be
further improved, as discussed above.

A particular problem exists in the $|V_{ub}|$ determination, namely a
very large $B\to X_c l\nu$ background. Various kinematic cuts can be used
to suppress the background.
The fractions of events with the different cuts are listed in Table 1.

\begin{table}
\caption{Kinematic cuts for suppressing the $B\to X_c l\nu$ background.}
\begin{tabular}{|c|c|} \hline\hline
Kinematic cut & Fraction of events\\ \hline
$E_l>(M_B^2-M_D^2)/(2M_B)$ & $\sim 10\%$\\ \hline
$q^2>(M_B-M_D)^2$ & $\sim 20\%$\\ \hline
$M_X<M_D$ & $\sim 80\%$\\ \hline\hline
\end{tabular}
\end{table} 
Imposing kinematic cuts causes additional theoretical uncertainties,
since the extrapolation in phase space to obtain the desired observable 
involves
theory. The hadronic invariant 
mass ($M_X$) cut is most efficient. It retains a vast majority of
phase space, so that only a small extrapolation to the entire phase space
is needed. The other kinematic cuts are less efficient and would cause
larger theoretical uncertainties because of a considerable dependence on
dynamic subtlety.

\section{A NEW METHOD FOR THE PRECISE DETERMINATION OF $|V_{ub}|$}

In the light-cone limit, $y^2\to 0$, the $b$-quark number conservation
leads to the sum rule \cite{new,sum}
\begin{equation}
S\equiv\int_0^1 d\xi_u\, \frac{1}{\xi_u^5}
\frac{d\Gamma}{d\xi_u} 
=|V_{ub}|^2\frac{G_F^2M_B^5}{192\pi^3}
\label{eq:sumrule}
\end{equation}
for the charmless inclusive semileptonic decay $B\to X_u l\nu$.
The sum rule establishes a theoretically clean relationship between the
observable $S$ and $|V_{ub}|$. The sum rule has the following advantages:
(1) it is model-independent (in particular, independent of the
distribution function), (2) it has no reliance on quark-hadron duality,
(3) no free parameters (such as $m_b$) except for $|V_{ub}|$ enter the
sum rule, and (4) it receives no perturbative QCD corrections. Therefore,
by using the sum rule, a more reliable and precise value of $|V_{ub}|$ 
can be determined than by using the conventional method.

Of course, to obtain the theory-motivated observable $S$ in $B\to X_u
l\nu$ one must deal with the large $B\to X_c l\nu$ background. The
strategy appears to be:
(1) apply the kinematic cut $M_X<M_D$ (or other realistic cuts), (2)
measure the weighted spectrum $\xi_u^{-5}d\Gamma/d\xi_u$, (3) extrapolate
it to entire phase space to obtain the integral $S$, and (4) determine
$|V_{ub}|$ from the observable $S$ by using the theoretically clean
relationship (the sum rule) between them.

This method for determining $|V_{ub}|$ is analogous to the determination
of $|V_{cb}|$ from the observable, $d\Gamma/dw$ at zero recoil,
in the exclusive semileptonic decay $B\to D^* l\nu$. Both use a
theoretically clean relationship, which is derived from QCD symmetries,
between the observable and $|V_{ub}|$ or $|V_{cb}|$.
Both require an extrapolation in phase space to obtain the
theory-motivated observable.

There are two sources of theoretical uncertainties in this method for the
$|V_{ub}|$ determination, one from higher twist corrections to the sum
rule and another from the extrapolation to obtain $S$.
In principle, the light-cone expansion systematically takes into account
higher twist corrections.
The quantitative estimate using the heavy quark effective theory has
shown that the error on $|V_{ub}|$ from higher twist corrections to the
sum rule amounts to around $1\%$ \cite{twist}.
The error on $|V_{ub}|$ due to the shape of the distribution function
used for the extrapolation has been assessed at the $6\%$ level \cite{sum}. 
With improved knowledge of the shape of the distribution function, as
discussed before, one could reduce this error.

\section*{ACKNOWLEDGMENTS}

This work was supported by the Australian Research Council.

\end{document}